# Micellization Thermodynamic Behavior of Gemini Cationic Surfactants. Modeling its Adsorption at Air/Water Interface

Vicente Domínguez-Arca[1*], Juan Sabín[1], Pablo Taboada[2], Luís García-Río[3], Gerardo Prieto[1*]

[1]Biophysics and Interfaces Group, Department of Applied Physics, Faculty of Physics, University of Santiago de Compostela, 15782 Santiago de Compostela, Spain

[2]Grupo de Física de Coloides y Polímeros, Departamento de Física de la Materia Condensada; Universidad de Santiago de Compostela, 15782-Santiago de Compostela, Spain.

[3]Centro Singular de Investigación en Química Biolóxica e Materiais Moleculares (CIQUS), Departamento de Química Física, Universidade de Santiago, 15782 Santiago, Spain

*Author to whom correspondence should be addressed
Tel.: +34 881814039
Fax: +34 881814112
Email: xerardo.prieto@usc.es , vicente.dominguez@rai.usc.es

## ABSTRACT

Self-assembly structures of gemini surfactants are characterized, among others, for their low *CMC*. This characteristic could be due to great hydrophobic parts in their molecular structures. That availability could imply great stability of self-assembly structures or monolayers absorbed in an interface.

The micellization behavior of two cationic gemini surfactants, α,ω-bis(S-alkyl dimethylammonium) alkane bromides, were studied by a modelization of dynamic surface tension (DST) experimental data and isothermal titration calorimetry (ITC) measurements. The adsorption data at the air/water interface was taken through the analysis of the profile changes of a pendant drop. The thermodynamic characterization of the micellization process of the gemini surfactants was carried out using ITC.

A model based on the Frumkin adsorption isotherm and the Ward-Tordai diffusion equation was developed to obtain the characteristic parameters of the adsorption without the need of using the Gibbs adsorption equation. Positive values of lateral interaction show good stability of the adsorbed monolayer. The ITC data were analyzed following a novel protocol based on the identification of the different energetic contributions and regimens observed in the titration enthalpograms from demicellization processes. The presence of exothermic peaks would explain the low values of *CMC*.

# 1. INTRODUCTION

The use of amphiphilic molecules is of crucial importance in many different industries such as chemical, oil recovery, medicine or personal care ones[1, 2]. The main characteristic of these compounds is their exotic capability of self-assembling in aqueous solution under certain thermodynamic conditions. Great efforts have been performed in order to explain and model their self-assembly behavior and adsorption processes at interfaces[3-5], which are based fundamentally on entropic interactions, Van der Waals forces and hydrogen bonding. The better performance and higher efficacy of surfactant molecules is directly related to their ability to decrease their critical micelle concentration and surface tension as well as to enhance their emulsification and solubilization properties, which can be exploited for the introduction of innovative and more effective surfactant-based products. In this regard, Gemini surfactants appear as a relatively new class of innovative surfactant molecules which can fulfill this need. These are dimeric surfactants that consist of two monomeric head groups connected by a spacer[6-8]. In comparison with their corresponding monomeric counterparts, gemini surfactants show a higher surface activity, and lower critical micelle concentrations (*CMCs*) and Kraft temperatures[9, 10], as well as forming in solution many different morphological micellar structures, vesicles, helices or tubules[11, 12]. All these properties make them suitable for many different industrial processes[13, 14] as emulsifiers, dispersants, coating agents and corrosion inhibitors, and biophysical applications[2, 15, 16] such as membrane solubilization, drug delivery systems or gene delivery by DNA-compaction, amongst others.

In the present work, the self-assembly behavior of two new gemini surfactants 1,4-Bis(tetradecyl trimethyl ammonium) butane (14-4-14) and 1,6-Bis(tetradecyl trimethyl ammonium) hexane was perforemd. Apart from the previously mentioned applications, this kind of surfactants bearing quaternary ammonium salts have also shown, for example, excellent properties as microbiocides[17], permeation enhancers of drugs[16, 18], biodetereoration inhibitors[19] or as cleaning and disinfectant agents[20]. Also, the potential influence of changes in the spacer length on the resulting physic-chemical properties were considered. Dynamic surface tension (DST) and isothermal titration calorimetry measurements (ITC) were done in order to characterize the newly synthesized surfactants, and new models to manage surface tension and calorimetric data and explained the obtained results were developed.

In this regard, it is worth mentioning that the variation of the surface tension with the surfactant bulk concentration is clasically analyzed using the Gibbs adsorption theory. However, experimental observations show incompatibilities between the saturation behavior at the interface of cationic surfactants and the assumption of this saturation regime before the surfactant CMC according to the Gibbs adsorption theory[21]. Regard to the saturation behavior, there is a high depth discussion germinated from several concatenated works based on questions and answers between reputed researchers in the field[22-26]. The Gibbs adsorption equation:

$$\Gamma = \frac{-1}{nRT}\left(\frac{d\gamma}{dlnc_b}\right)\Big|_{T,P}$$

(1)

is applied for experimental data obtained at constant temperature and pressure, where $n$ is the dimensionless constant accounting for the surfactant ionic state, $R$ is the ideal gas constant ($J\,K^{-1}mol^{-1}$), $T$ the temperature ($K$), $\gamma$ is the measured surface tension ($N\,m^{-1}$), $c_b$ is the surfactant bulk concentration ($M$), and $\Gamma$ is the surface excess concentration ($mol\,m^{-2}$). It is usually accepted that $n = 1$ for non-ionic and $n = 2$ for ionic surfactants, respectively.

However, for gemini surfactants the value of $n$ remains doubtful due to the ionic structure of these surfactants. The use of Eq. (1) has been previously questioned in the literature[22], so there is a clear need to develop new models to explain the origin of surface tension changes[27]. Then, we here developed a new method to analyze the dynamic adsorption isotherm of gemini surfactants based on the dynamic surface tension profiles at different bulk surfactant concentrations. This model is based on the Frumkin adsorption isotherm[28] to take into account the lateral chain interactions at the interface, consequence of the ionic character and large hydrophobic moiety of the chosen surfactants (tails and spacer). In this regard, the ionic nature of the present surfactants leads to electrostatic interactions between the adsorbed gemini molecules, so a parameter to take into account such lateral interactions is introduced ($A$). The large hydrophobic moieties interact via Van der Waals attractive forces, so the interface behaves as attractive ($A > 0$) and the adsorption is cooperative. On the other hand, the problem of diffusion from bulk to the solution interface was treated through the Ward-Tordai diffusion equation[29]. The relationship between the dynamic surface tension and the surfactant concentration was established by means of the May-Jeelani-Hartland empiric equation[30]. Then, our approach allow us to explain and model the behavior of the dynamic surface tension profiles at the beginning of adsorption process, in which experimental data show an initial inflection only compatible with a cooperative adsorption isotherm.

On the other hand, demicellization ITC data from the derive gemini surfactants were shown to display both endothermic and exothermic peaks upon each titration (at surfactant concentrations in the sample cell below the *CMC*) for this kind of gemini surfactants. Because that exothermic contribution disappears in the micellar dilution regime (gemini concentration in the sample cell > *CMC*) a novel method is also implemented to get an explanation about the exothermic contribution in the demicellization regime, and a new protocol based on this exothermic component was used to estimate the *CMC*.

# 3. MATERIALS AND METHODS

## 3.1. Materials

The gemini surfactants 1,4-Bis(tetradecyl trimethyl ammonium) butane (14-4-14) and 1,6-Bis(tetradecyl trimethyl ammonium) hexane were synthesized from the corresponding $\alpha,\alpha'$-dibromide, 1,4-dibromobutane or 1,6-dibromohexane (5 mmol), and anhydrous N,N-dimethyltetradecylamine (10 mmol) in 50 mL of acetone upon boiling under reflux for 96h. The material obtained after removal of the solvent with a rotary evaporator was crystallized from ethanol-ether. The obtained crystals (recrystallized from methanol) were, then, dried in a vacuum desiccator at ambient temperature to give the desired product (25% yield). All other reactants were of the highest commercially available purity and were used as received.

Both surfactants were characterized by $^1$H NMR spectroscopy (see Figure SI1). NMR spectra were measured in $D_2O$ in a Bruker Advance ARX-400 spectrometer operating at 400 and 100 MHz, respectively. The observed $^1$H NMR shifts (300 MHz, $D_2O$, 25°C) for 14-4-14 (Figure SI1A) were: $\delta$=3.50 (m, 8H), $\delta$=3.22 (s, 12H), $\delta$=1.95 (m, 4H), $\delta$=1.82 (m, 4H), $\delta$=1.34 (m, 44H), $\delta$=0.92 (m, 6H) and for 14-6-14 (Figure SI1B): $\delta$=3.41 (m, 8H), $\delta$=3.19 (s, 12H), $\delta$=1.82 (m, 8H), $\delta$=1.34 (m, 48H), $\delta$=0.93 (m, 6H).

### 3.2. Pressure-controlled pendant-drop surface balance

A previously described pendant drop tensiometer was used[31, 32]. Briefly, this instrument uses Axisymmetric Drop Shape Analysis (ADSA)[33] to determine the surface tension, contact angle, drop surface area and drop volume based on a drop profile. A camera records drop profiles at a determined frequency; at the same time, a controllable injection device is responsible to keep constant the drop volume or surface. The drop is generated and pended at the end of a micro-polished tip. The process of drop formation starts with an injection of a determined volume of a surfactant solution of known concentration. Immediately, the software starts to record and analyze the drop profiles to calculate the dynamic surface tension. The drop is inside a glass cuvette semi-coated with a thermostatic flow and water vapor to guarantee isothermal measurements at the liquid-vapor equilibrium. In our experiments, the drop volume was kept constant to ensure the same surfactant concentration during the diffusion process to the drop surface.

### 3.3. Isothermal Titration Calorimetry

The titrations were performed using an isothermal titration microcalorimeter (VP-ITC) from Microcal Inc. (Northampton, MA) at constant pressure. In each run, a solution of 3 mM gemini surfactant ($C_{syr}^{gem}$) in a 0.270 mL automatic syringe was sequentially injected (5 μL each injection) into the sample cell containing water, while stirring at 100 rpm. The concentration of gemini surfactant inside the sample cell ($C_{cell}^{gem}$) is increasing upon each titration. Samples and water were degassed and thermostated by using a ThermoVac accessory before the titration process. The equilibration time between injections was set at 300 s.

## 4. RESULTS AND DISCUSSION

### 4.1. Thermodynamics of micellization of gemini surfactants

Isothermal microcalorimetry was used to analyze the interactions involved in the micellization behavior of the present gemini surfactants. It is typical to find in the literature procedures to analyze the titration enthalpograms of surfactant demicellization based on the determination of an inflection point in the heat plot by sigmoidal regression of the demicellization enthalpies[34, 35] to determine the *CMC*. In a similar procedure, the demicellization processes at different temperatures[36] or different media were also analyzed[37]. Typical demicellization enthalpograms of gemini surfactants are given in Figure 1. All these curves show similar profiles, with two clearly identified sections in each plot. When $C_{cell}^{gem}$ > *CMC* (right side of raw data), the titration enthalpogram is always composed of endothermic peaks. Within this concentration range, the micelle is a thermodynamically stable structure, so the heat flow can be attributed to only a micelle dilution phenomenon. This hypothesis is consistent with the fact that the enthalpy decreases as the concentration increases as a consequence of the rearrangement of the micellar system after titration, where dilution is less energetic as the concentration of micelles becomes higher. When $C_{cell}^{gem} < CMC$ (left side of raw data) the depicted process is more complex. Experimental data show the coexistence of both endothermic and exothermic peaks after each titration as well as a strong temperature dependence. An intuitive and novel method is developed in this work to analyze the most relevant energetic components at $C_{cell}^{gem} < CMC$. Titrations of the left side show a demicellization process composed of three different phenomena: i) a micelle-breakup (MB) represented by the energy necessary to separate micelles in gemini surfactants; ii) a second phenomenon when the micelle was broken and the hydrocarbon parts of gemini surfactants –chain and spacer groups- become in contact with water molecules, leading to the appearance of caged water molecules, named here as dress-water formation (DW); and iii) finally, an energetic contribution related to surfactant dilution which appears when micelles are broken and monomeric surfactants are surrounded by water molecules. This latter phenomenon will be neglected in our methodology since it is considered the smallest energetic contribution.

In enthalpic terms, the micelle-breakup (MB) process is associated with the presence of endothermic peaks. The increase in the separation between surfactant molecules causes a gain in potential energy, which involves a decrease in sample cell temperature which is compensated by heat flux from reference to sample cell, so that $\Delta H_{MB} > 0$. On the other hand, the dress-water formation (DW) process is associated with exothermic peaks and can be represented as a binding phenomenon, then $\Delta H_{DW} < 0$. In this case, water molecules form thermodynamic stable structures or clusters around the surfactant hydrocarbon moieties decreasing their freedom. The outcome is a sample cell temperature increase due to the solvent molecules not involved in those clusters, so an exothermic process occurs. This could be observed in any demicellization process for any kind of surfactant. In the case of gemini, the exothermic peaks are stronger as a consequence of their larger hydrophobic surfaces within their molecular structure caged with water molecules with hydrogen bonding.

Both MB and DW processes are depicted in the first part of the left side of raw data shown in Figure 1. The right side of raw data shows the dilution regime of surfactant micelles since the concentration is high enough to ensure their stability. Between the two commented regimes -dilution of micelles and MB/DW processes- a smooth transition is also observed.

In Figure 2, enthalpies resulting from the integration exclusively of the observed exothermic peaks are shown. These obtained profiles show the DW process ($\Delta H_{DW}$) upon changes of surfactant concentration seem to a binding interaction. When micelle-breakup takes place, a rearrangement of water molecules immediately forms a layer (or "dress", caged-water model) around the hydrocarbon surfactant chains. Thus, a binding model is applied to derive thermodynamic information from these exothermic peaks. In this model, the substrate would be the "water dresses" and the ligand the surfactant hydrocarbon chains. This method gives us new criterium to determine the *CMC* since when the exothermic component vanishes the substrate is saturated. So, all "water dresses" are used and only a process is possible when the concentration increase: the micellization.

In this way, given a system formed by a certain ligand concentration ($[L]_T$) and substrate ($[S]_T$) we can apply a single set of identical sites (SSIS) binding model[38] using a mass balance:

$$[L]_T = [L] + \iota[S]_T \tag{2}$$

where $[L]$ is the concentration of free ligand while the binding is taking place, and $\iota$ is the average occupation. When $\iota = 0$, there is no binding and when $\iota = 1$ each substrate hole or "water dress" is occupied. We could consider a binding isotherm assuming the existence of no preferred interactions between the ligand and the substrate, this is, a Hill-Langmuir isotherm:

$$K = \frac{\theta}{(1 - \theta)[L]} \tag{3}$$

where $K$ is the adsorption or equilibrium constant, and $\theta$ is the occupation fraction of the substrate by the ligand. When $\theta = 0$ the substrate is fully empty, and if $\theta = 1$ the substrate is fully bonded. There is a relation between $\theta$ and $\iota$ given by:

$$\theta = \frac{\iota}{N} \tag{4}$$

where $N$ is the occupancy at each substrate. If $N = 1$, the occupation fraction and average occupation are the same.

To fit plots in Figure 2, a relationship between the evolved heat ($Q$) and the enthalpy ($\Delta H_{DW}$) is proposed:

$$Q = N\theta[S]_T \Delta H_{DW} V_0 \qquad\qquad (5)$$

where $V_0$ is the volume of the solution in the sample cell. The model applied to fit the data is the Single-Site Model from Origin 7.0 with Microcal LLC ITC script. Briefly, the model is same that showed in eq. (5) with an implemented correction in the volume due to successive titrations.

Table 1 displays the obtained quantities after fitting the experimental values by application of the developed binding model to $\Delta H_{DW}$ values. The *CMC* values were estimated from the point at which the exothermic contribution ceased, denoting a typical parabolic behavior with temperature. The values of the binding constant ($K_{DW}$) are compatible with a favorable adsorption process. The Gibbs free energy of dress water formation, $\Delta G_{DW}$, was negative for both gemini surfactants indicating the spontaneity of the process, and are similar as a result of their small structural differences.

In Figure 3a, changes in the micellization enthalpy values with temperature are displayed. The micellization enthalpy will be of the same magnitude but of different sign to that of demicellization enthalpy. This enthalpy is estimated by summing the areas from the endothermic to the exothermic peaks in each titration, taking into account the net sign of the involved energy. The procedure is shown in Figure 3b, in which each titration is composed of an endothermic (micelle break up) and an exothermic peak (resulting from the dress water formation around the hydrocarbon chains). To estimate the demicellization enthalpy, only the average of the five first titrations was considered by assuming that other further titrations display additional energy contributions, such as energies involved in premicellar states or the decrease of enthalpy in MB process.

The micellization Gibbs free energy was estimated from the *CMC* values shown in Table 1 and the following equation for dimeric surfactants bearing monovalent counterions[34, 39]:

$$\Delta G_{mic}^{chain} = RT \left( \frac{1}{2} + \beta \right) ln(cmc) - \left( \frac{RT}{2} \right) ln(2) \tag{6}$$

In eq. (6), the *CMC* is expressed in terms of molarity of each alkyl chain with data from Table 1, and $\beta$ is the fraction of charges of micellized univalent surfactant ions neutralized by micelle-bound univalent counterions. This parameter is calculated by conductivity measurements using $\beta = 1 - \alpha$ and literature data for structurally-related $C_{14}TAB$[40] surfactant. The estimated values of $\Delta G_{mic}$ can be observed in Figure 3a. As it can see, the values of $\Delta G_{mic}$ lay into a narrow range for all temperatures. The negative sign in all cases indicates that the micellization is a spontaneous process. Once obtained $\Delta G_{mic}$ and $\Delta H_{mic}$, it is possible to calculate $\Delta S_{mic}$ using the Gibbs free energy for a system under an isobaric and isothermic process:

$$\Delta G_{mic} = \Delta H_{mic} - T \Delta S_{mic} \tag{7}$$

Figure 3a shows the changes of $\Delta G_{mic}$, $\Delta H_{mic}$ and $\Delta S_{mic}$ with solution temperature. As previously reported for similar systems[36] at low temperatures $\Delta H_{mic} > 0$ and $\Delta S_{mic} > 0$, and the micellization process is entropy-driven. As the solution temperature decreases, the motion of water molecules becomes more restricted, that is, the number of configurations decreases. Also, the hydrocarbon surfactant chains induce a decrease in the freedom of water molecules in their surroundings, and when the number of surfactant molecules is large enough, a rearrangement of the hydrocarbon chains forming micelles occurs to increase the freedom of those water molecules, which is characteristic of an entropically-driven process. As the water temperature increases, the hydrogen bonding strength between water molecules surrounding alkyl chains decreases, the motion of water molecules is faster and the micellization process is governed by van der Waals interaction between the surfactant alkyl groups. Thus, our measurements of the surfactant demicellization at high temperatures mainly represent the energy involved in the break of surfactant micelles due to the screening of van der Waals interaction originated from the thermal agitation of water molecules, which implies that $\Delta H_{mic} < 0$ and $\Delta S_{mic} < 0$ being characteristic of an enthalpic-driven process.

Figure 4 shows the relation between $\Delta H_{mic}$ and $\Delta S_{mic}$, which can be envisaged in Figure 3a. This relation might represent an example of an enthalpy-entropy compensation process, experimental evidence reported and discussed in several adsorption experiments and demicellization process[41-44]. Although there is a great amount of experimental data that shows its presence in numerous systems[45, 46], that compensation is a not yet fully explained effect. The free energy $\Delta G$ plays the main role in the enthalpy-entropy compensation. If the sign of $\Delta G$ is negative and the $\Delta H$ decrease with temperature, the sign of the slope in the compensation line is negative. Figure 3a shows a narrow range of variation of $\Delta G_{mic}$ for all temperatures, while the $\Delta H_{mic}$ observed decreases in a huge range when the temperature increase. In some reported experiments, the compensation effect in narrow ranges of $\Delta G$ was studied from the first law of thermodynamics[47]. So the intercept in the compensation line shows the free energy and the slope a compensation temperature, which is near to the experimental temperature. The data reported in this work shows enthalpies and free energies took at different temperatures of the demicellization process, as a previously reported and discussed topic[48], the compensation and the parameters of the line can not be explained using the same ways that systems in which the free energy window is narrow and the temperature is kept constant. Taking account that free energy is what forces the enthalpy-entropy compensation, the nature of the intercept and slope of the compensation line will be evaluated.

In Figure 4, the region with $\Delta S_{mic} > 0$ corresponds to an entropically-driven micellization and other with $\Delta H_{mic} < 0$ corresponds to an enthalpy-driven one. A potential linear relationship between them can be inferred through:

$$\Delta H_{mic} = \Delta H^* + T_c \Delta S_{mic} \tag{8}$$

where $\Delta H^*$ is the micellization enthalpy corresponding to $\Delta S_{mic} = 0$. As this process is reversible (isentropic), the energy of micellization will be entirely considered as the chemical part of the process. Therefore, $\Delta H^*$ represents the surfactant-surfactant interaction[35]. The slope $T_c$ is a compensation temperature. To understand the meaning of this temperature, it is recommendable to rewrite eq. (8) as:

$$\Delta S_{mic} = \frac{\Delta H_{mic} - \Delta H^*}{T_c} \tag{9}$$

It is worth mentioning that two interactions are fundamental in a surfactants micellization process: solute-solute and solute-solvent interactions. Solute-solvent interactions might be characterized by $T_c$ since the numerator in eq. (9) provides the energy resulting from the desolvation process, $\Delta H_{mic}$ quantifies the energy corresponding to micellization process and $\Delta H^*$ corresponding to solute-solute interactions. The magnitude of $T_c$ will determine, then, the entropy variation for the specific micellization process. As can be seen in Figure 4, there are no large differences between gemini surfactants. These small changes resulted from the small structural differences in their molecular structure.

## 4.2. Surfactant adsorption at the a/w interface from dynamic surface tension (DST) data

A theoretical model was developed to analyze the dynamic adsorption of gemini surfactants from the water bulk solution to the air/water (a/w) interface. Several similar models and procedures to deal with this process can be found in the literature[27, 42, 49-52], but the present proposed dynamic model allows derive the necessary quantities for the equilibrium state. A model based on a set of three equations taking into account diffusion, adsorption at the interface and the relationship between concentration in the neighborhood between the interface and surface is developed to make use of dynamic surface tension data. This procedure has several advantages as, for example, the possibility to analyze not only the adsorption isotherm of equilibrium values but also different dynamic adsorption isotherms. In this manner, we will obtain values of maximum concentration at the interface at any bulk concentration, adsorption constants, and lateral interaction constants but, in this case, using a *classical* adsorption process. In this model, there are depicted three different zones, which play different roles in the adsorption process, as seen in Figure 5. The bulk area is the zone in which the surfactants are dispersed or also self-assembled in the form of micelles depending on the initial solution concentration. The subsurface is defined as a plane located just below the interface, and which represents a part of the bulk solution that undergoes an important concentration change as a result of the adsorption process. This plane has a surfactant concentration ($c_s(t)$) lower than the bulk concentration and can change to satisfy the dynamic equilibrium with the solute adsorbed at the interface. Finally, the adsorption plane or interface is an area where the surfactant is adsorbed and in which this phenomenon is assumed to be instantaneous in the proposed model.

In a general diffusion theory allowing back-diffusion, the dynamic mass transfer between the bulk and the subsurface can be modeled by the Ward-Tordai equation[29]:

$$\Gamma(t) = 2c_b\sqrt{\frac{D}{\pi}}\sqrt{t} + 2\sqrt{\frac{D}{\pi}}\int_0^t c_s(\tau)d(\sqrt{t-\tau}) \tag{10}$$

where $D$ is the diffusion coefficient, $c_b$ is the bulk concentration, $\tau$ is a variable of integration, $\Gamma(t)$ is dynamic surface concentration and $c_s(t)$ is the time-depending subsurface plane concentration. As developed, eq. (10) is basically the result of applying the Fick's laws to a non-homogeneous solution of a diffusion problem. The first term corresponds to a homogeneous solution, whilst the second one represents a component due to a variable condition, $c_s(t)$.

Since we assume instantaneous adsorption from the subsurface plane to the interface, we will here present the model that reproduces how our gemini surfactants are adsorbed at the interface. Besides, we will consider the lateral chain and polar head interactions in the adsorption process. In summary, the adsorbed concentrations $\Gamma(t)$ and $c_s(t)$ are related by the Frumkin´s adsorption isotherm[28, 53]:

$$c_s(t) = \frac{1}{K} \frac{\Gamma(t)}{\Gamma_m - \Gamma(t)} e^{-A\frac{\Gamma(t)}{\Gamma_m}} \tag{11}$$

where $K$ is an adsorption constant in terms of the relative interfacial concentration $\frac{\Gamma(t)}{\Gamma_m}$, $\Gamma_m$ is a maximum surface excess concentration for a determined $c_b$, and $A$ is a parameter related to lateral interactions. Both Eq. (10) and Eq. (11) can be solved simultaneously taking into account boundary conditions and initializing parameters. Using experimental data of dynamic surface tension isotherms ($\gamma(t)$), it is necessary to establish a relationship between the concentrations $\Gamma(t)$, $c_s(t)$ and $\gamma(t)$. To do that, it is necessary to discern firstly whether surface tension is a *bulk* or *surface* phenomenon. There are not few efforts involved to conclude the role of bulk concentration and surface coverage in the surface tension phenomenon. For the present authors, that question is key at the start point of discussions between the applicability[23, 54, 55] or not[21, 22, 25] of the Gibbs adsorption isotherm in the study of surface tension variations by surfactant activity. In the powerful thermodynamic derivation of the Gibbs adsorption isotherm lies a huge handicap to apply this formalism to surfactant species with not-conventional ionic structure, such as gemini surfactants, which activity to be necessary to know and to estimate an adequate $n$ in Eq. (1). On another hand, as a fundamental model, the Gibbs adsorption isotherm in general terms predict a variation of surface tension due to the increase of bulk concentration after full coverage of the surface occurred. Due to this, using Eq. (1) should be enough to estimate a $\Gamma_m$ by determination of slope in a $\gamma_{eq}$ $vs$ $\log(C_b)$ plot if hypotheses of Eq. (1) could be fully accepted for these kinds of surfactants.

If the separation between two phases is analyzed such as an air/water interface, it is possible to envisage surface tension as a property related to the *cohesion* of water molecules. In this way, when the interaction between the molecules of a medium increases, the surface tension also does, denoting surface tension like a *bulk* phenomenon. On the other hand, if at the interface there is a third species different than air or water, the surface tension will vary with respect to the a/w interface. Hence, we can show the surface tension like a property related to the concentration of a material adsorbed or in the neighborhood of the interface, denoting surface tension like a *surface* phenomenon. As a conclusion, at the air/water interface in which the surfactants are adsorbed, we here introduce a variational form for the surface tension depending on the surfactant concentration near the interface:

$$\left.\frac{d\gamma(t)}{dc_s(t)}\right|_{T,P,c_b} = -\kappa(\gamma(t) - \gamma_*) \tag{12}$$

where $\gamma(t)$ is the time-dependent surface tension, $c_s(t)$ is the time-dependent subsurface surfactant concentration, $\kappa$ is an adsorption constant in terms of the relative surface tension, and $\gamma_*$ is the surface tension at equilibrium for $c_b$. The variational form of Eq. (12) shows that the surface tension decreases more slowly when the value of $\gamma(t) - \gamma_*$ is increasingly smaller. However, there is a limit for the decrease of the surface tension, $\gamma_*$. This form relates implicitly the surface tension variation to the interface surfactant adsorption and subsurface filling. In this regard, there exists a maximum occupancy in the subsurface to satisfy the dynamic equilibrium with the surfactant present within the bulk solution. As seen in Figure SI2, all values of bulk concentration satisfy $c_s(t \to \infty) < c_b$.

Integrating Eq. (12) allow us to obtain a functional form to use the obtained experimental dynamic surface tension data:

$$\gamma(t) = \gamma_* + (\gamma_0 - \gamma_*)e^{-\kappa c_s(t)} \tag{13}$$

where $\gamma_0$ is the surface tension of water. Eq. (13) is similar as the empirical May-Jeelani-Hartland equation[30] proposed to analyze the behavior of surface tension equilibrium values upon changes of bulk concentration. From experimental data, $\gamma(t)|_{c_b}$, we obtain estimated values for $\Gamma_m$, $A$, $\gamma_*$, $K$ and $\kappa$. A numerical computation routine was developed to solve simultaneously Eqs. (10), (11) and (13). This function returns a specific $\gamma$ value for a given $t$.

Typical DST plots are given in Figures 6-7. These show profiles denoting changes in the surface tension from an initial value to an equilibrium one depending on the bulk surfactant concentration. The model assumes that the surface tension variation is directly related to the sub-surface concentration of gemini surfactant, $c_s(t)$ -Eqs. (12) and (13)- and indirectly to the surface coverage –eq. (11)-. Experimental data and fits in Figures 6-7 allowed to estimate different adsorption parameters, which are shown in Table 2. It is necessary to comment that the used diffusion coefficients for gemini surfactants *14-4-14* and *14-6-14* were assumed to be similar[56, 57] $D = 3.00 \times 10^{-10} \, m^2/s$. This assumption is based on the presence of small structural differences between both surfactants, which hardly affects their diffusion coefficients. *CMC* values obtained by the classical determination from equilibrium surface tension data are also depicted in Figures 6-7. As seen, values of $0.18mM$ and $0.19mM$ for *14-4-14* and *14-6-14* gemini, respectively, were obtained and are in agreement with those obtained from ITC data (see Table 1).

Table 2 displays the obtained values for $\kappa$, $K$, $A$ and $\Gamma_m$ after the fitting of experimental data to the developed model. Both adsorption constants $\kappa$ and $K$ have large values compatible with the evidence that these gemini surfactants possess *CMC* values several orders of magnitude lower than the structurally-related single chain surfactants. As the gemini surfactants have larger hydrophobic moieties than their homologous single counterparts, the hydrophobic effect is more important for the former and, thus, *CMC* values are much smaller.

As mentioned previously, Eq. (11) introduces a parameter $A$ through the Frumkin isotherm. Only positive values of $A$ were able to reproduce the initial surface dynamic tension changes (see Figure 8), denoting the great stability of the adsorption surfactant layer as a consequence of the attractive lateral interactions between adsorbed gemini molecules compared to a Langmuir adsorption[58] ($A = 0$). As the interface tends to saturation, Eq. (12) was proposed to explain the surface tension variation. These changes are caused not only by the surface saturation but also by the increase in the equilibrium values of $c_s$ as $c_b$ increases. This behavior can be observed in Figure SI2, where $c_s$ values at equilibrium is smaller than their respective $c_b$ values.

Values of maximum surface excess concentrations, $\Gamma_m$, are shown in Table 2. As seen, $\Gamma_m$ is larger for *14-4-14* surfactant than for *14-6-14* one as a consequence of the larger size of the latter molecule and its higher hydrophobicity. As the used diffusion coefficients were the same for both gemini, the relation $\Gamma_m^4 > \Gamma_m^6$ can be observed from the fact that the dynamic isotherms of *14-6-14* decay before than the isotherm of *14-4-14* at the same concentration (see Figure 8). In other words, the equilibrium surface tension is reached before for *14-6-14* than for *14-4-14* at similar concentrations. This phenomenon can be explained using the qualitative results applying a classical analysis based in Gibbs adsorption. From Eq. (1), $\Gamma_m$ is directly related to the slope of the logarithmic plot of $\gamma_{eq}$ vs. $c_b$.

## 5. CONCLUSIONS

Following previously reported procedures with some slight modifications[59] two new (α,ω-bis(S-alkyl dimethylammonium) alkane bromide gemini surfactants with different spacer length were obtained, and their surface activity and thermodynamics of micellization studied by means of dynamic surface tension and isothermal titration calorimetry measurements.

Firstly, the equilibrium and kinetic behaviors of the cationic *gemini* surfactants at the air/water interface was analysed using the pendant drop technique. The derived experimental data were explained in terms of a new model based on the Frumkin adsorption isotherm and the Ward-Tordai diffusion, which allowed us to explain the surfactants surface active behavior at the initial stages of the dynamic adsorption process at the air-water interface in terms of a cooperative adsorption process, in contrast to many previous works[21, 22, 27, 50, 60,] [61]. In addition, positive values of lateral interactions indicate a high stability of the surfactant monolayer formed at the interface[62-65].

The proposed model allows us to avoid the use of Gibbs adsorption isotherm to not enter in conflict with the hypothesis about the chemical potential of the gemini surfactant, arguments in agreeing with previous works[21, 22, 25]. Nevertheless, the presented model shows the compatibility between a full surface covered and the variation of surface tension by the increase of bulk concentration, arguments in agreeing with classical methodologies[55] and other critic works with the fact to avoid Gibbs adsorption isotherm[23, 54].

On the other hand, ITC demicellization data at different temperatures revealed the coexistence of endothermic and exothermic peaks upon each titration in the micelle breakup regime. A new intuitive method was also developed to analyze the two energetic components of the obtained titration peaks in this regime, and to determine the extremely low *CMC* values of these kinds of surfactants, which also showed a strong temperature dependence. This new methodology allows the decoupling of hydration and micelle breakup contributions considering the former as a binding process, which is an improvement over previously developed analysis ITC[34-36, 41, 44].

Future work will be directed to study and to characterize the stability of monolayers and bilayers formed by lipids and cationic gemini surfactants to exploit their properties in biomedical applications including drug entrapment/release, biomembrane modelling or functionalization of biointerfaces as well as to develop new models or refined existing ones in order to get information about the stability of micelles or aggregation number from thermodynamic data.

**Acknowledgments:** This work was supported by Ministerio de Economia y Competitividad of Spain (projects MAT2016-80266-R and CTQ2017-84354-P), Xunta de Galicia (Grupo de Referencia Competitiva ED431C 2018/26; Agrupación Estratégica en Materiales-AEMAT ED431E 2018/08; GR 2007/085; IN607C 2016/03 and Centro singular de investigación de Galicia accreditation 2016–2019, ED431G/09) and the European Union (European Regional Development Fund-ERDF), is gratefully acknowledged.

**Table 1** Thermodynamic parameters derived from the ITC calorimetry and the application of the SSIS model.

| | $T$ (K) | $\Delta H_{DW}$* ($kJ \cdot mol^{-1}$) | $\Delta G_{DW}$ ($kJ \cdot mol^{-1}$) | $T \cdot \Delta S_{DW}$* ($kJ \cdot mol^{-1}$) | $K_{DW}$* ($M^{-1}$) | $CMC$** ($M$) |
|---|---|---|---|---|---|---|
| **14-4-14** | 278.15 | -5.01 | -33.16 | 28.15 | $1.72 \cdot 10^6$ | 0.23 |
| | 298.15 | -2.86 | -34.40 | 31.54 | $1.06 \cdot 10^6$ | 0.19 |
| | 303.15 | -2.37 | -35.20 | 32.83 | $1.17 \cdot 10^6$ | 0.23 |
| | 308.15 | -2.81 | -36.58 | 33.76 | $1.58 \cdot 10^6$ | 0.25 |
| **14-6-14** | 278.15 | -8.55 | -32.21 | 23.66 | $1.05 \cdot 10^6$ | 0.22 |
| | 280.65 | -4.94 | -32.87 | 27.93 | $1.33 \cdot 10^6$ | 0.20 |
| | 288.15 | -2.08 | -30.27 | 28.20 | $3.15 \cdot 10^5$ | 0.19 |
| | 293.15 | -2.36 | -33.01 | 30.65 | $7.80 \cdot 10^5$ | 0.19 |
| | 298.15 | -2.73 | -33.65 | 30.92 | $7.89 \cdot 10^5$ | 0.20 |
| | 303.15 | -3.48 | -29.09 | 25.61 | $1.29 \cdot 10^5$ | 0.22 |

*Results from apply Single Set of Identical Sites (SSIS) model. Dress-water formation around the hydrophobic sectors of the *gemini* surfactans could be understood as a binding phenomenon.

**CMC values were estimated by examination of the raw data. When there are not presence of exothermic enthalpy it was suppose that there are not dress-water formation, so the micelle are stable structures.

**Table 2:** Estimated values of the main parameters in the theoretical model to dynamic surface tension: $\kappa$ is a adsorption constant in terms of the relative surface tension; $\Gamma_m$ is a maximum surface excess concentration; $A$ is a parameter related with the lateral interactions; $K$ is an adsorption constant in terms of the relative interfacial concentration $\frac{\Gamma(t)}{\Gamma_m}$; $D$ is the diffusion coefficient.

|  | $\kappa^*$ $(M^{-1})$ | $K^*$ $(M^{-1})$ | $A^*$ | $\Gamma_m{}^*$ $(mol \cdot m^{-2})$ | $D^{**}$ $(m^2 \cdot s^{-1})$ |
|---|---|---|---|---|---|
| 14-4-14 | 123.80 | 724.38 | 1.16 | $7.40 \cdot 10^{-6}$ | $3.0 \cdot 10^{-10}$ |
| 14-6-14 | 140.35 | 816.46 | 0.93 | $6.31 \cdot 10^{-6}$ | $3.0 \cdot 10^{-10}$ |

*Results from apply the model propose in Section 2.

**Data for diffusion coefficient propose from bibliography.

## Figures Legends

FIGURE 1. Different parts identify in experimental demicellization raw data from ITC. Micelle dilution range is composed by endothermic peaks in the post-CMC regime. Endothermic peaks in pre-CMC regime are interpreted as the micelle-breakup process. The exothermic peaks in the pre-CMC regime at different temperatures are interpreted as dress-water (DW) formation process. The model proposed determines the CMC when the exothermic signals disappear. It shows that there is not DW formation process.

FIGURE 2. Experimental data from integration of exothermic peaks in raw titration data at 278.15 K for 14-4-14 (▲) and 14-6-14 (●). The smooth curve is the result from fitting experimental data to the SSIS binding model. Only two sets of data were shown, analogous process were carried out to elaborate Table 1.

FIGURE 3. a) Energy representation for enthalpy, entropy and free energy to micellization process. $\Delta H_{mic}$ was estimated by average of integration peaks of micelle-breakup and dress-water formation process. For averaging, just the five firsts titrations were took into account. b) Representation of areas per titration to estimate values of $\Delta H_{demic}$. The total area ($\Delta H_{demic}$) is calculated by addition of endothermic and exothermic areas taking into account their respective sign. Endothermic and exothermic peaks per titration in the demicellization regime were observed in all measurements. Pure endothermic peaks are only observed in the micellar dilution regime.

FIGURE 4. Enthalpy-entropy compensation plot at different temperatures. The linear regressions (Eq. 8) show estimated parameters $\Delta H^*$ and $T_c$.

FIGURE 5. Scheme representing the proposed diffusion model. Bulk, air/water interface and subsurface liquid plane. The adsorption at the interface from subsurface is assumed to occur instantaneously.

FIGURE 6. DST experimental data and fits applying the model at different concentrations for 14-4-14. A linear regression of pre-*CMC* equilibrium values was completed to compare the slopes between two species of gemini surfactants. The equation of fitted curve: $y = 20.19 - 25.20 \cdot log(c|b)$.

FIGURE 7. DST experimental data and fits applying the model at different concentrations for 14-6-14. A linear regression of pre-CMC equilibrium values was completed to compare the slopes between two species of gemini surfactants. The equation of fitted curve: $y = 24.06 - 22.46 \cdot log(c|b)$.

FIGURE 8. Comparison between DST curves at two different concentrations (0.030 mM and 0.050 mM) for 14-4-14 (○) and 14-6-14 (Δ) Gemini surfactants.

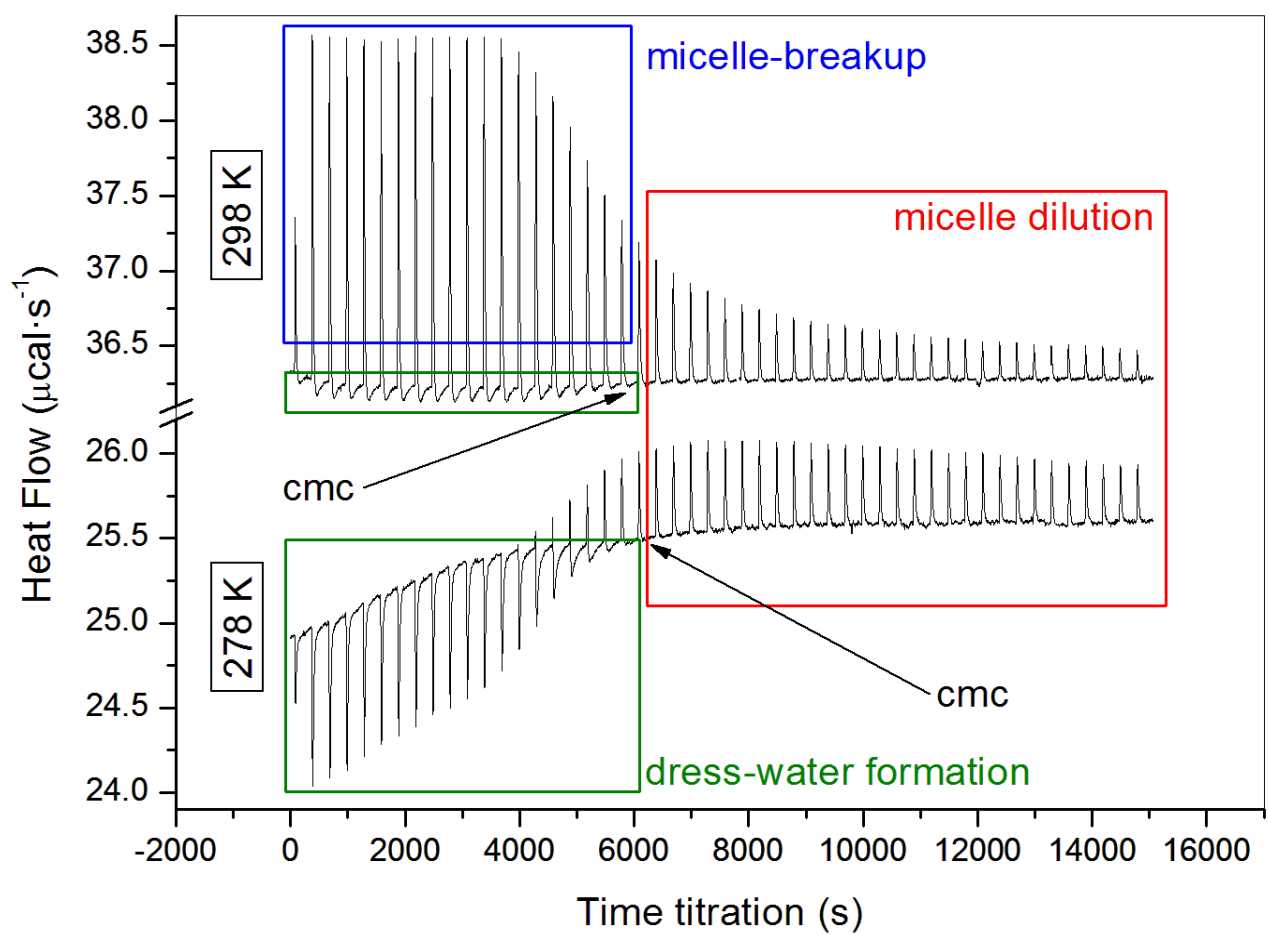

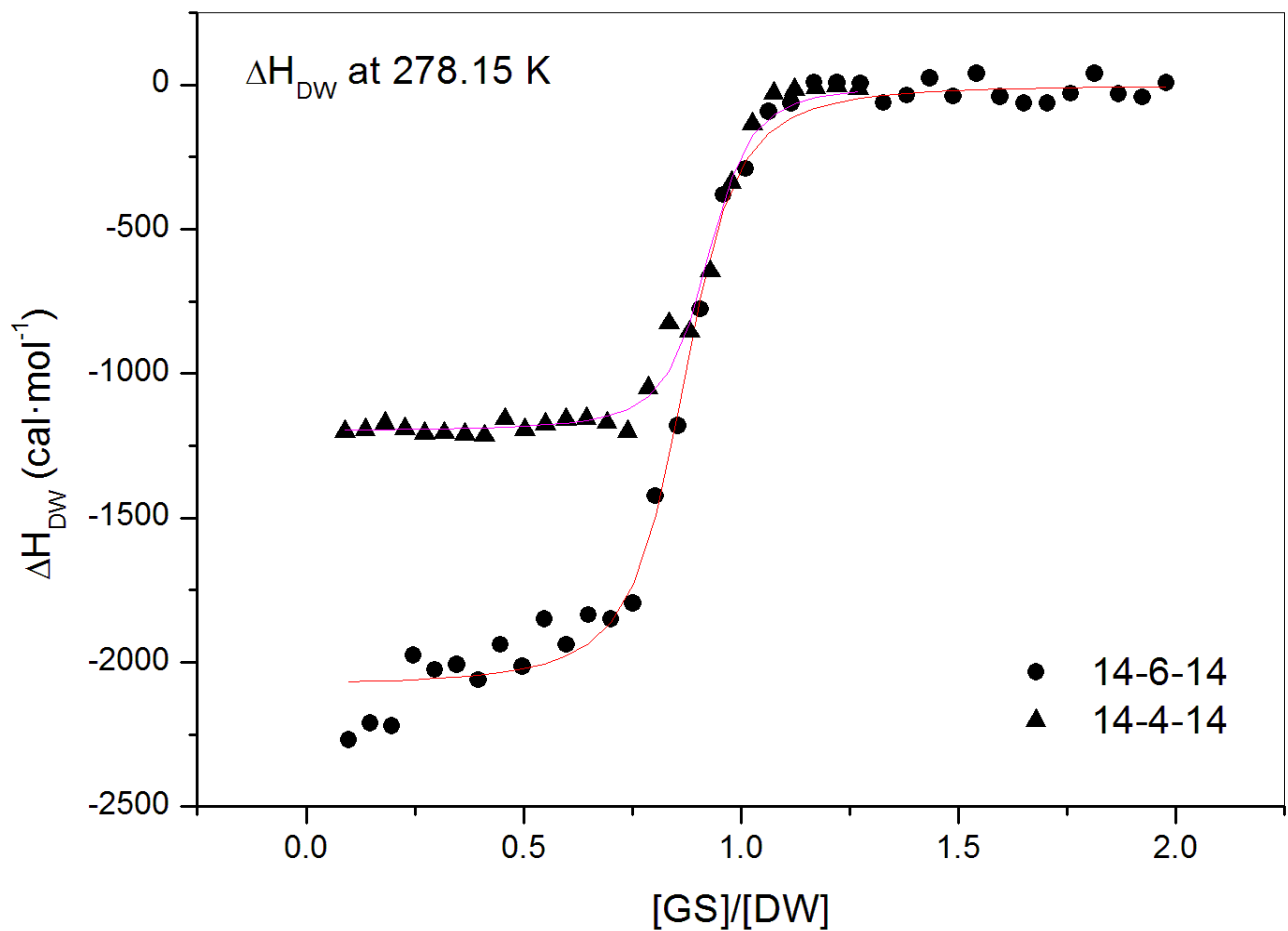

a)

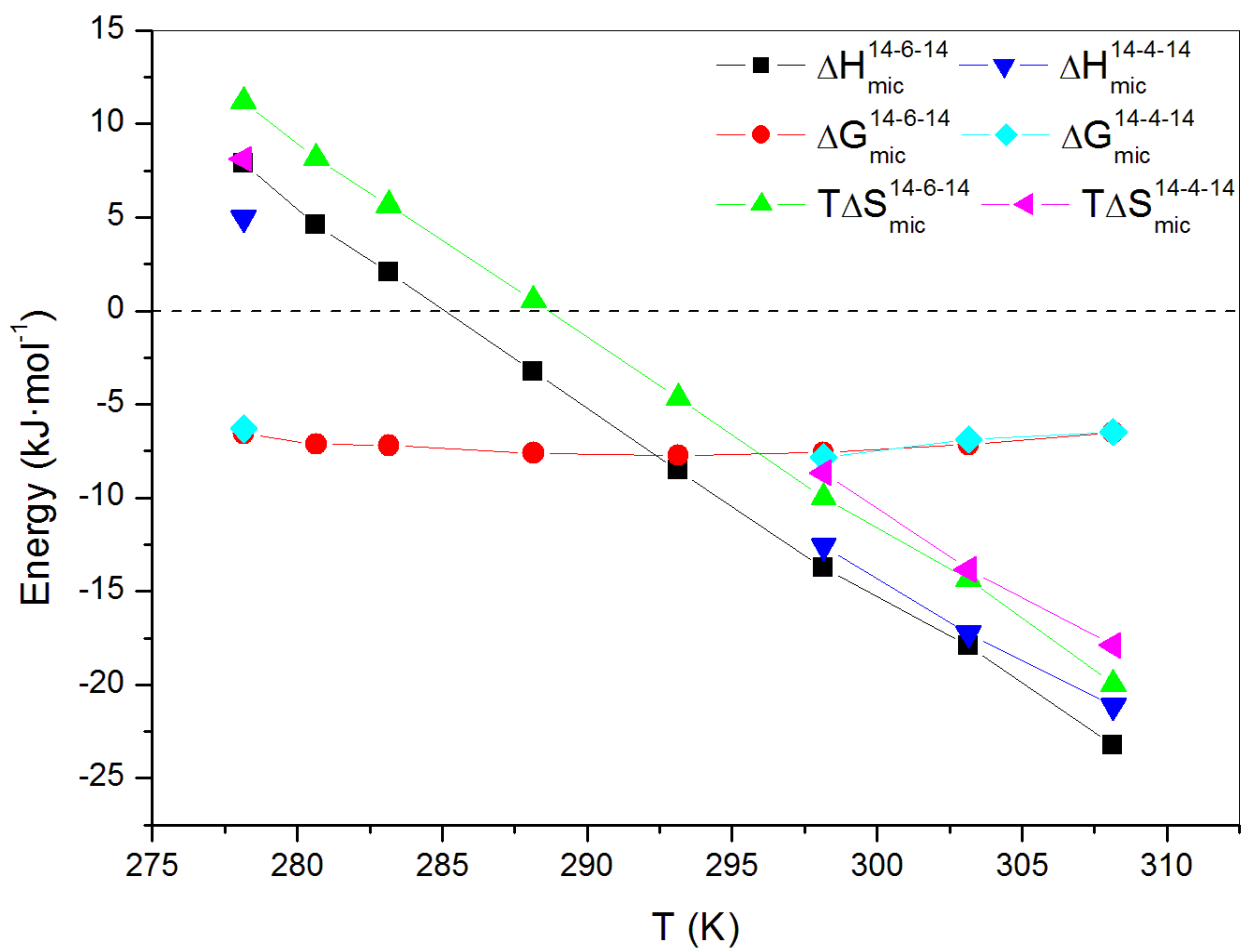

**b)**

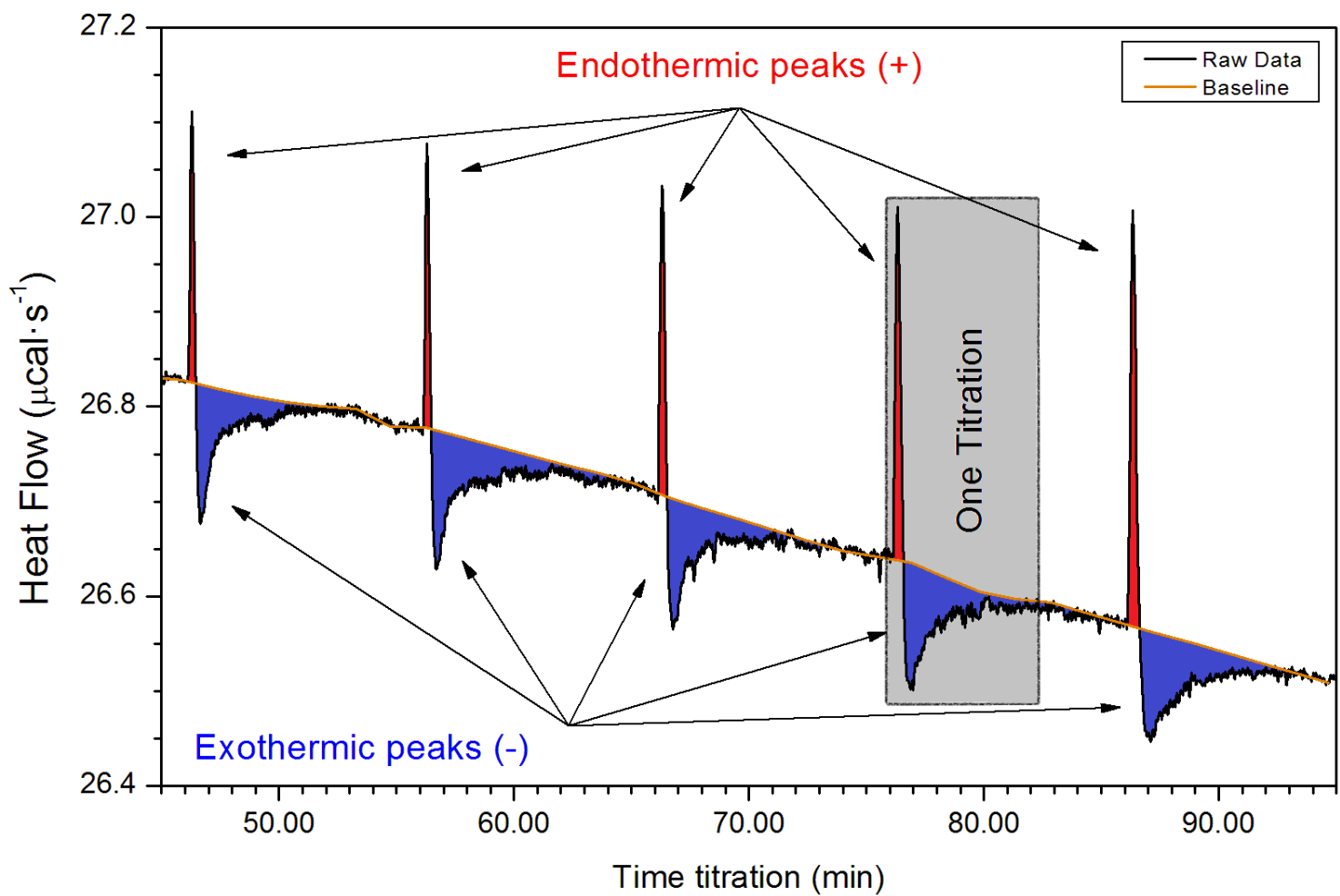

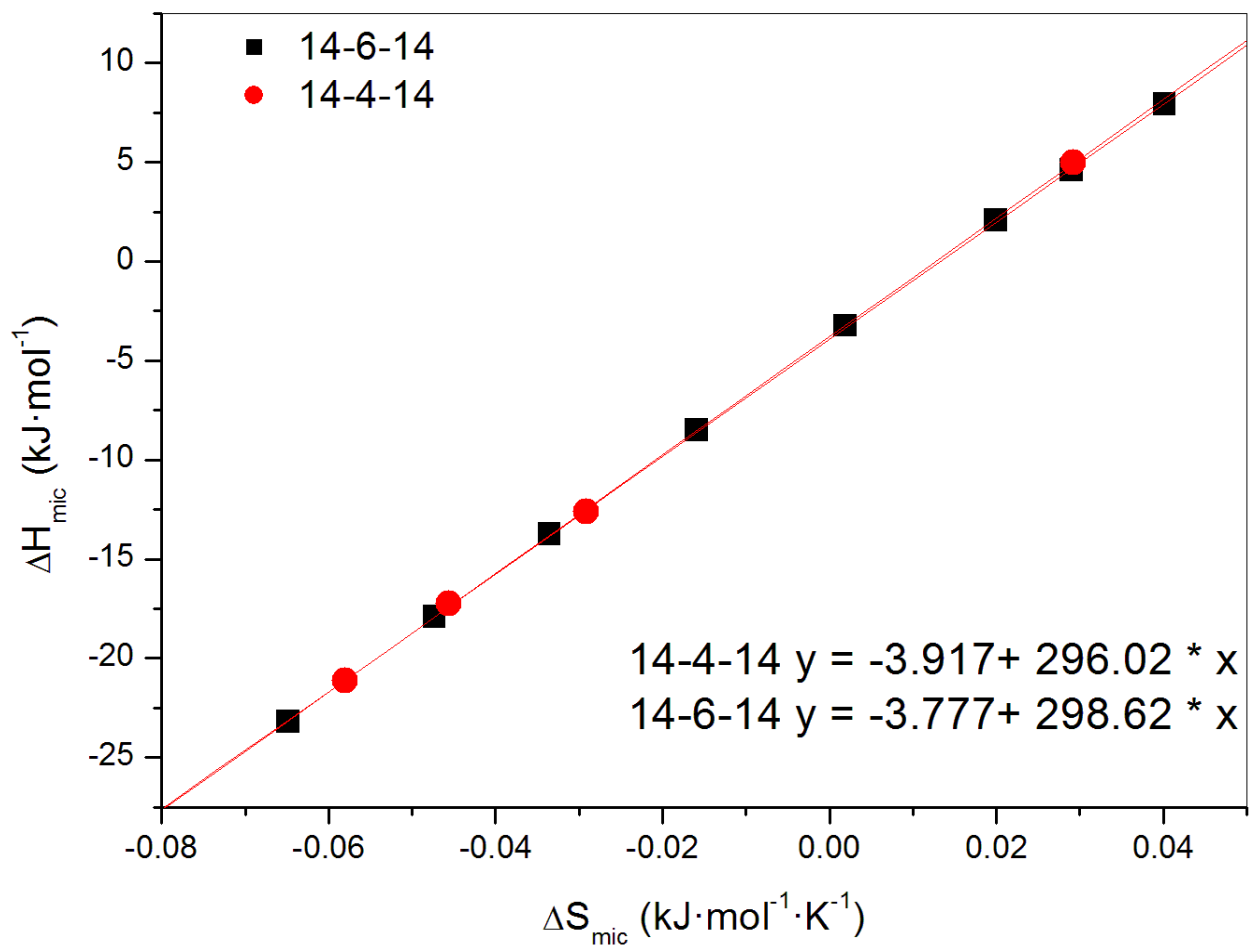

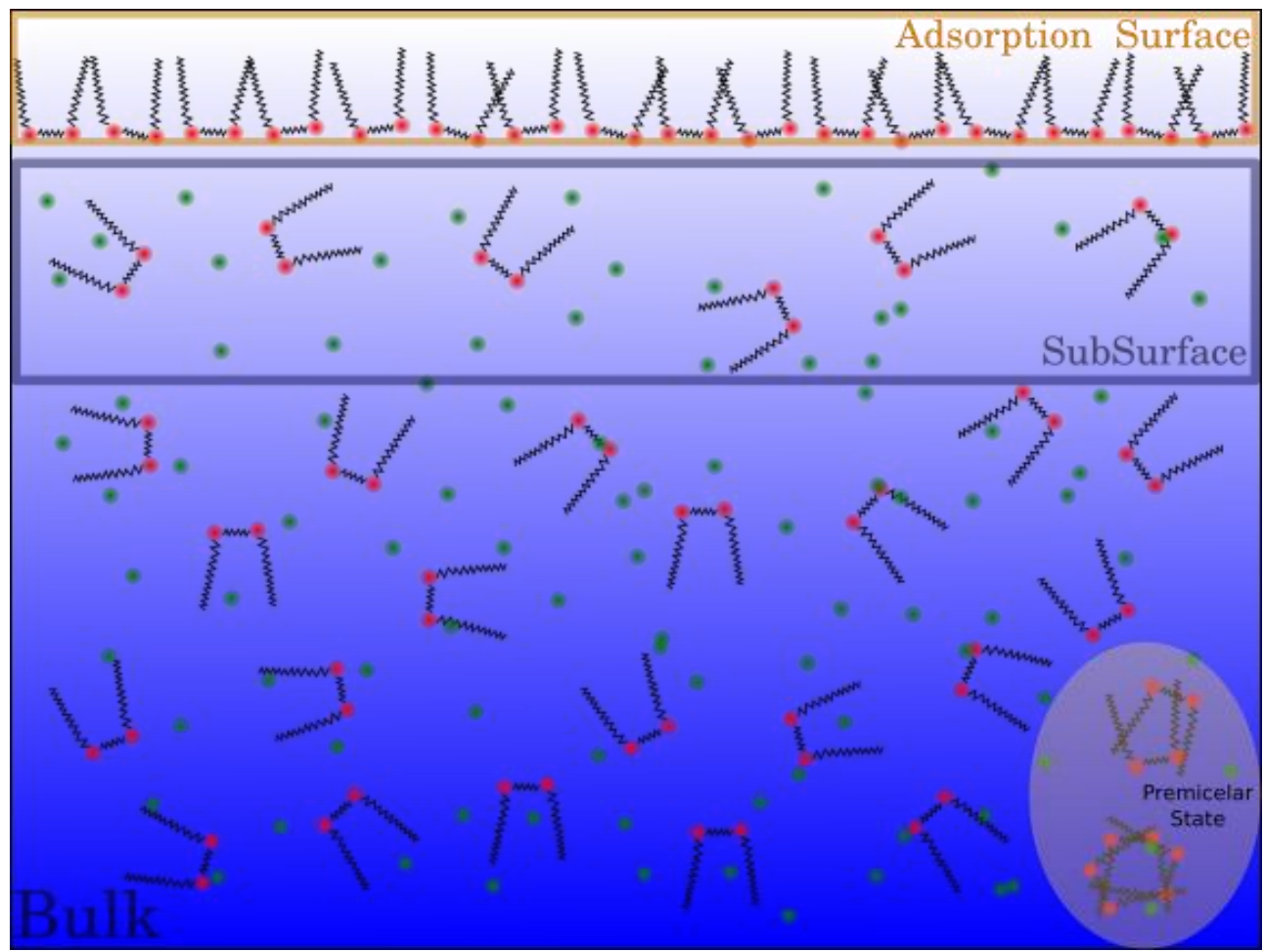

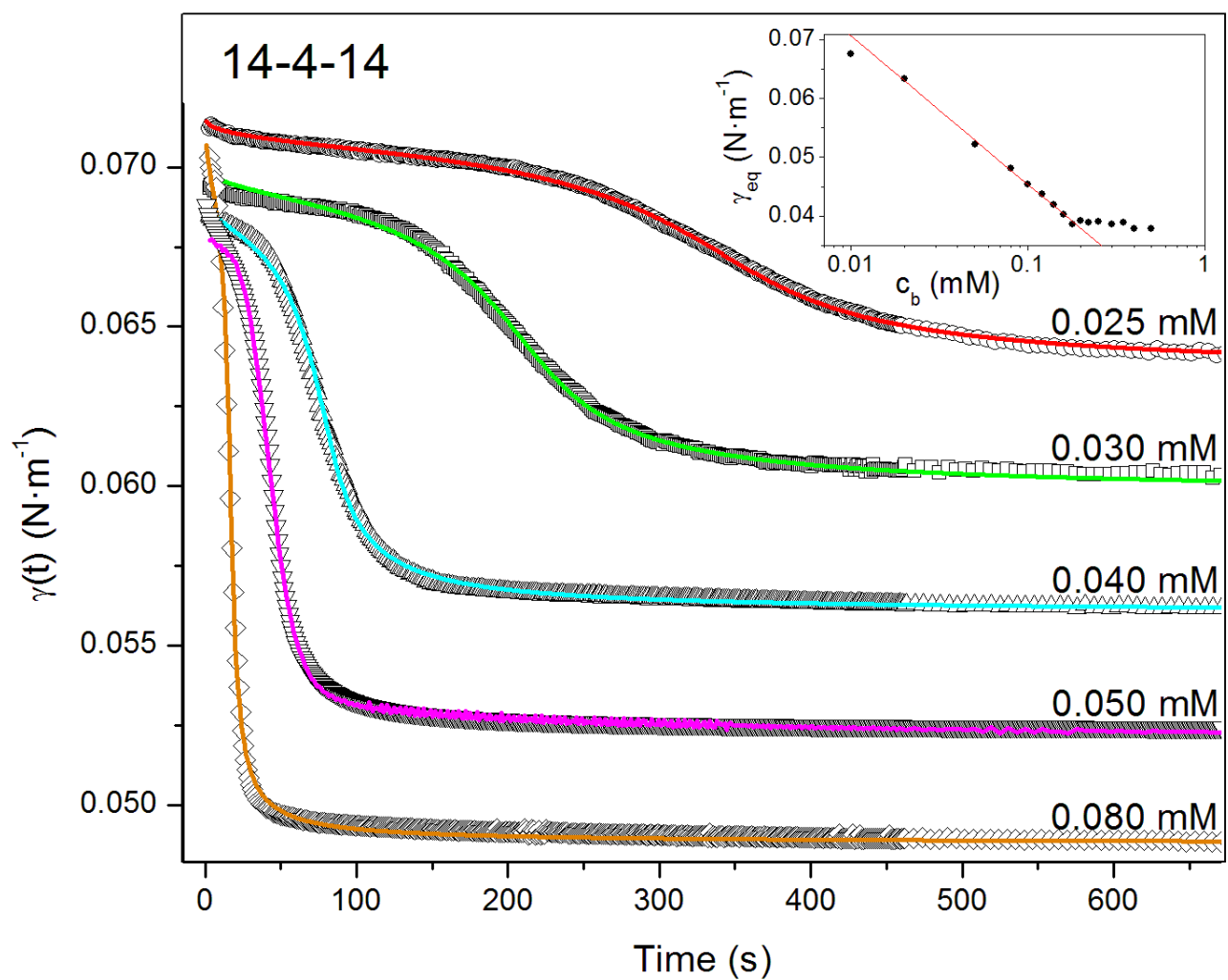

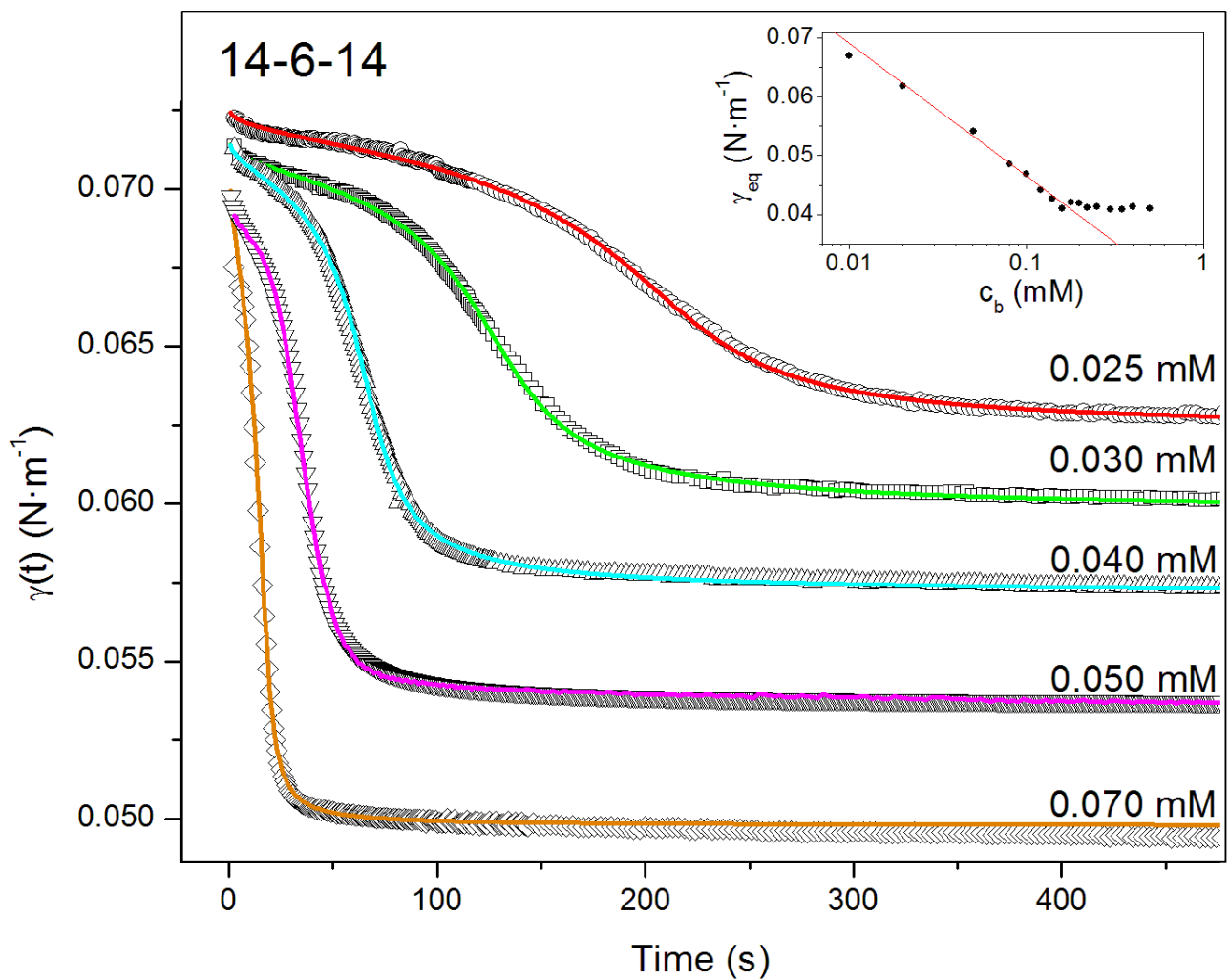

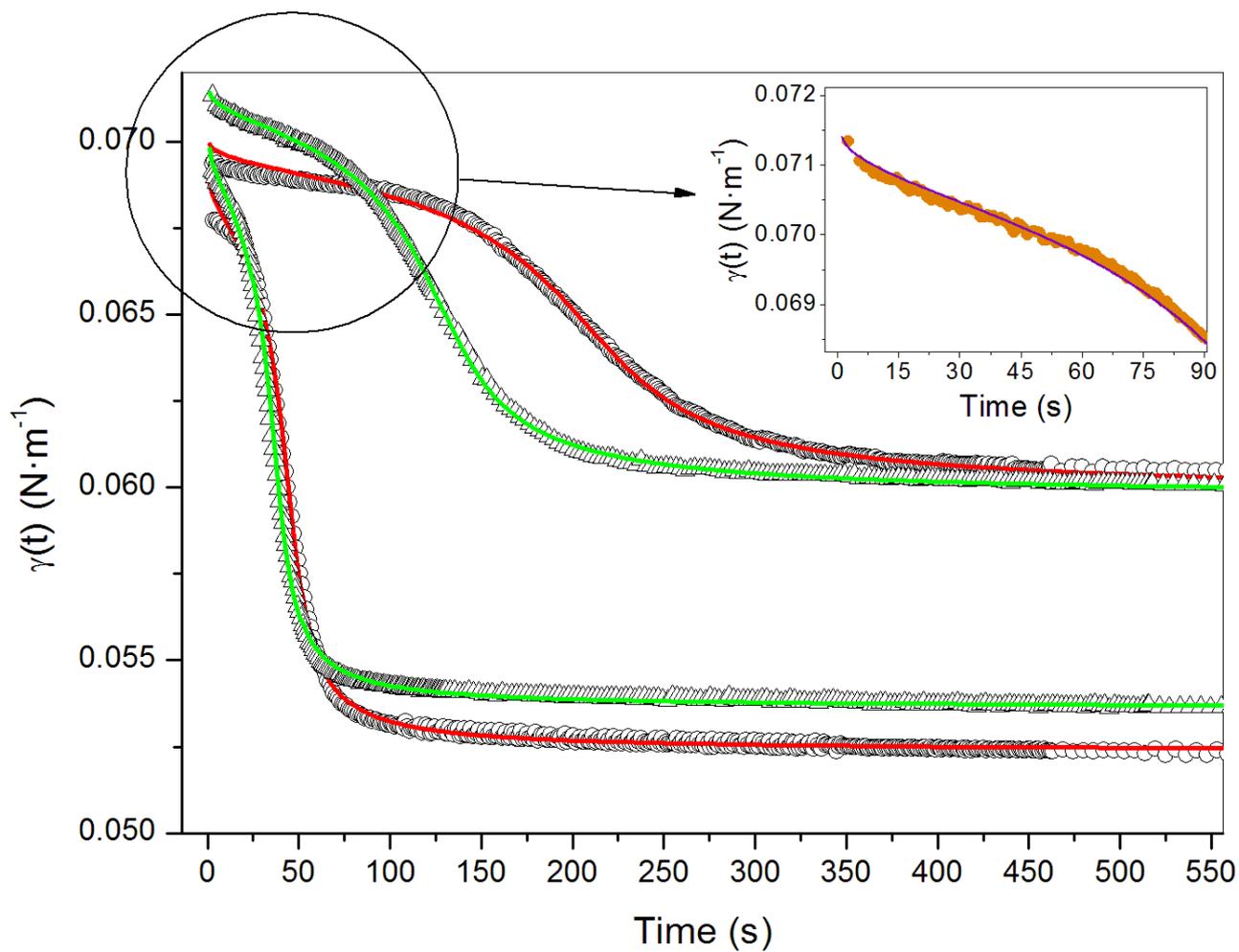

**Figure SI1**

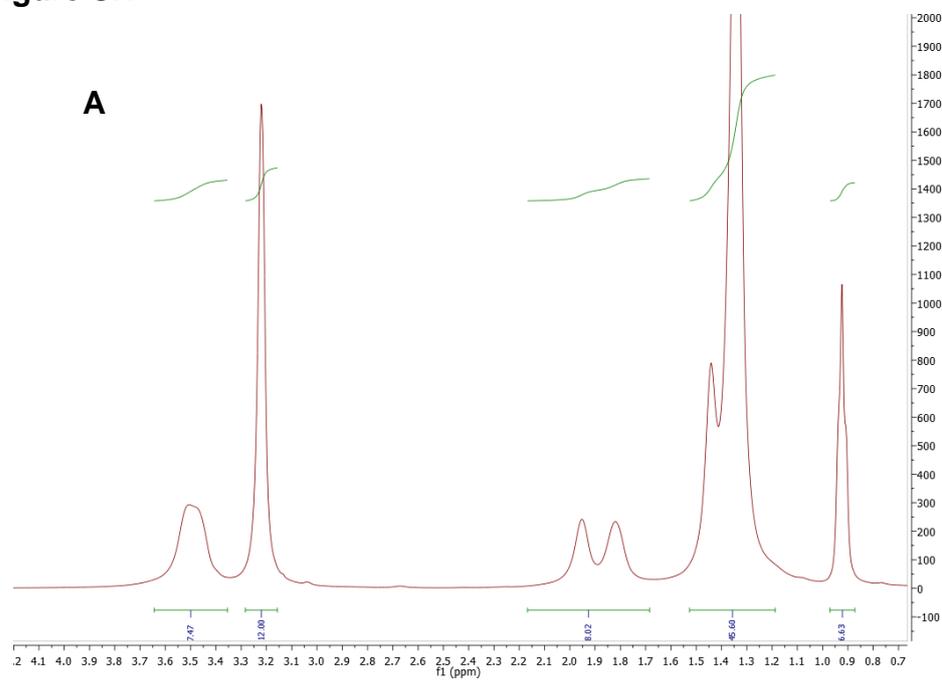

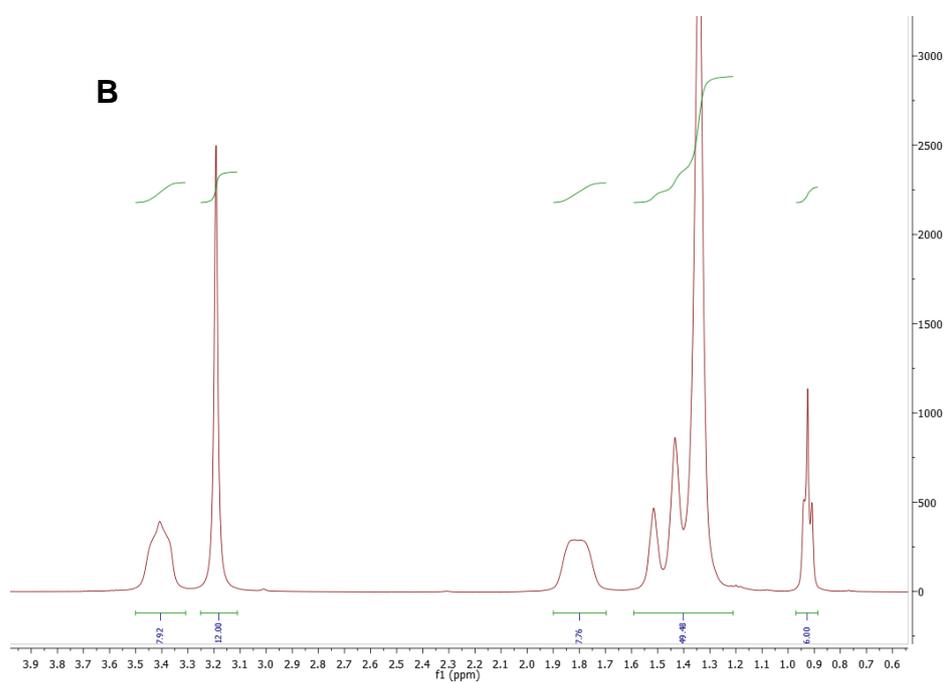

FIGURE SI1. $^1$H NMR spectrum (300 MHz in $D_2O$ at 25ºC) of A) 14-4-14 and

B) and 14-6-14.

**Figure SI2**

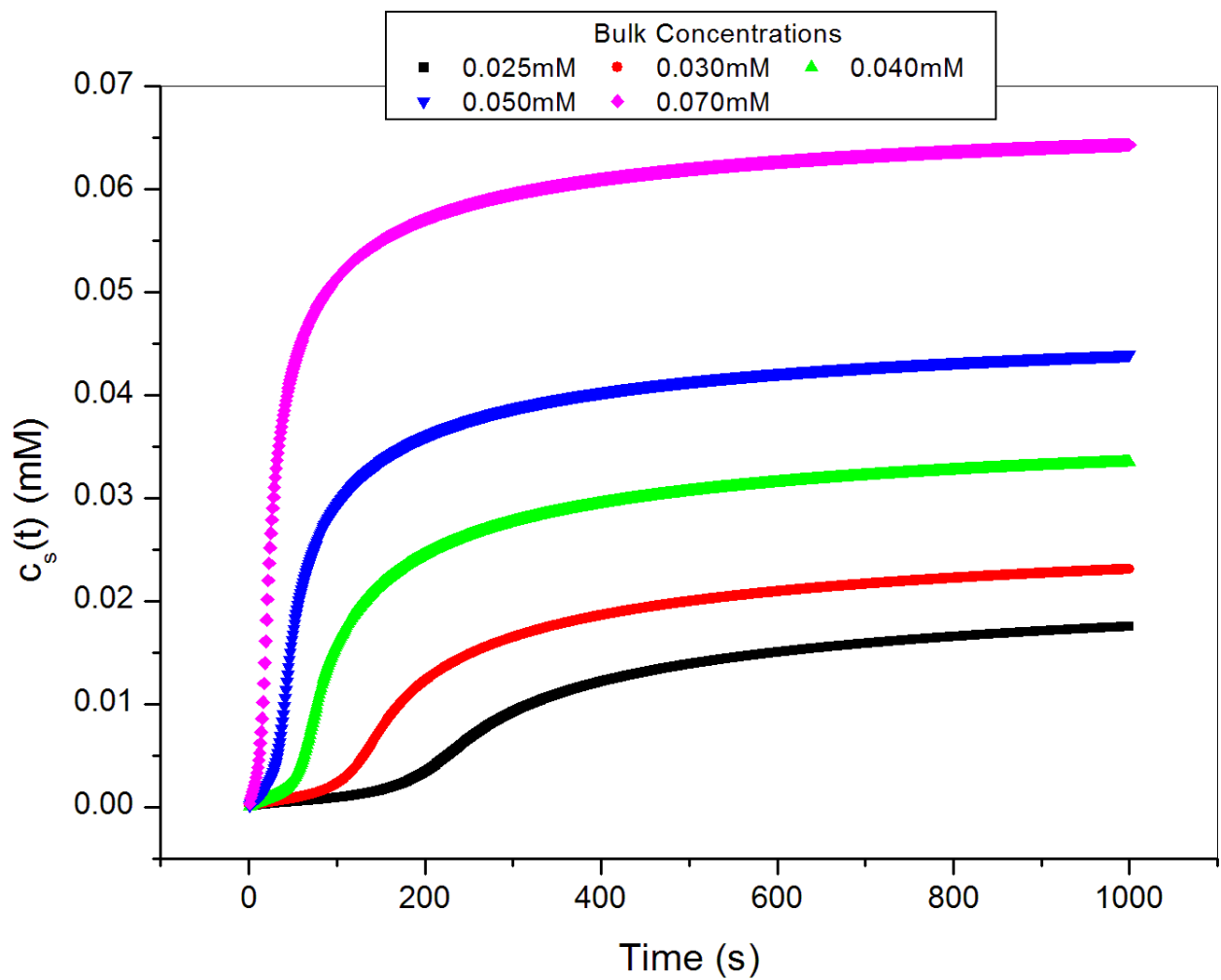

FIGURE SI2. Simulated temporal variation of the subsurface concentration with fitted values at different concentrations of 14-6-14 gemini surfactant.

**GRAPHICAL ABSTRACT**

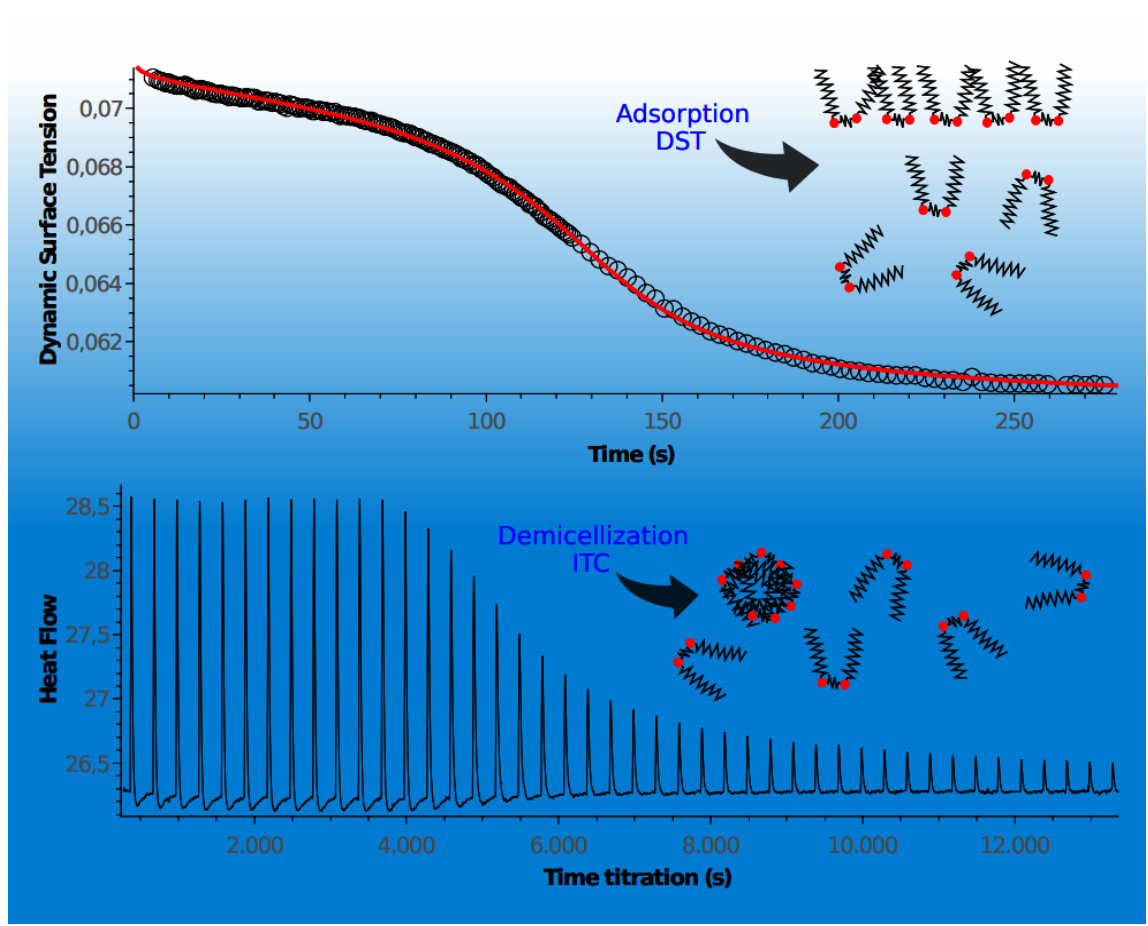